\documentclass[sigconf,authorversion]{acmart}

\usepackage{xspace}
\usepackage{xcolor}
\usepackage{enumitem}
\usepackage{graphicx}
\usepackage{fancyvrb}
\usepackage{amsmath}
\graphicspath{{graphics/}}
\usepackage{color}
\PassOptionsToPackage{hyphens}{url}\usepackage{hyperref}           
\usepackage{tabularx}
\usepackage{eucal}
\usepackage{booktabs}
\usepackage{amsfonts}
\usepackage{pifont}
\usepackage{subcaption}
\usepackage{float}

\hypersetup{
    colorlinks=true,
    linkcolor=blue,
    citecolor=blue,
    filecolor=red,      
    urlcolor=magenta,
    breaklinks=true,            
}

\usepackage{etoolbox}
\makeatletter
\patchcmd{\@makecaption}
  {\scshape}
  {}
  {}
  {}
\makeatletter
\patchcmd{\@makecaption}
  {\\}
  {:\ }
  {}
  {}
\makeatother

\newcommand{\cks}[1]{}
\newcommand{\cd}[1]{}

\newcommand{\tool}{\text{Isadora}\xspace} 

\usepackage{epstopdf}

\copyrightyear{2021}
\acmYear{2021}
\setcopyright{acmlicensed}\acmConference[ASHES '21]{Proceedings of the 5th Workshop on Attacks and Solutions in Hardware Security}{November 19, 2021}{Virtual Event, Republic of Korea}
\acmBooktitle{Proceedings of the 5th Workshop on Attacks and Solutions in Hardware Security (ASHES '21), November 19, 2021, Virtual Event, Republic of Korea}
\acmPrice{15.00}
\acmDOI{10.1145/3474376.3487286}
\acmISBN{978-1-4503-8662-3/21/11}

\begin{document}

\title{Isadora: Automated Information Flow Property Generation for Hardware Designs}
\author{Calvin Deutschbein}
\affiliation{%
  \institution{University of North Carolina at Chapel Hill}
  \country{}
}
\author{Andres Meza}
\affiliation{%
  \institution{UC San Diego}
  \country{}
}
\author{Francesco Restuccia}
\affiliation{%
  \institution{Scuola Superiore Santa-Anna Pisa}
  \country{}
}
\author{Ryan Kastner}
\affiliation{%
  \institution{UC San Diego}
  \country{}
}
\author{Cynthia Sturton}
\affiliation{%
  \institution{University of North Carolina at Chapel Hill}
  \country{}
}


\begin{abstract}
Isadora is a methodology for creating information flow specifications of
hardware designs.  The methodology combines information flow tracking and
specification mining to produce a set of information flow properties that are
suitable for use during the security validation process, and which support a
better understanding of the security posture of the design. Isadora is fully
automated; the user provides only the design under consideration and a testbench
and need not supply a threat model nor security specifications. We evaluate
Isadora on a RISC-V processor plus two designs related to SoC access
control. Isadora generates security properties that align with those suggested
by the Common Weakness Enumerations (CWEs), and in the case of the SoC designs,
align with the properties written
manually by security experts. \newline

\end{abstract}


\maketitle
\section{Introduction}

Security validation is an important yet challenging part of the hardware design
process. A strong validation provides assurance that the design  is
secure and trustworthy: it will not be vulnerable to attack once deployed, and it will
reliably provide software and firmware with the advertised security
features. A security validation engineer is tasked with defining the threat
model, specifying the relevant security properties, detecting any
violations of those properties, and assessing the consequences to
system security. 

Existing commercial design tools (e.g., Mentor Questa Secure 
Check, Cadence JasperGold Security Path Verification, and Tortuga Logic 
Radix) can verify security properties of a design, but the tools are only as
strong as the provided properties.
\emph{Defining these hardware security properties is a crucial part of the security validation process that currently involves a significant manual undertaking.}  We propose an automated methodology that combines information flow tracking with
specification mining to create a human-readable 
 information flow specification of a hardware design.
 The specification can be used as a set of security
properties suitable for use with existing security validation tools, and it can
also be studied directly by the designers to support their understanding of how
information flows through the design.

Information Flow Tracking (IFT) is a powerful security verification technique that monitors
 how information moves through a hardware design. Recently, IFT has been demonstrated at the RTL~\cite{Armaiti2017,Armaiti2018} and 
gate level~\cite{hu2014glift, Hu2016, Becker2017GLIFT}, and has been used to
monitor implicit flows through digital side
channels~\cite{bidmeshki2015toward, armaiti2017clepsydra,pilato2019tainthls}. 
Existing verification engines that incorporate IFT capabilities can be used to
confirm whether a given information flow property holds. However, it is up to the designer to specify the full set of
desired flow behaviors. 

The technique of specification mining 
offers an automatic
alternative to manually writing properties. 
Specification mining can be applied
to software~\cite{ammons2002mining} and hardware~\cite{hangal2005iodine}
and has recently been applied to system on a chip (SoC) 
designs~\cite{Rawat2020Hyperminer, Farzana2019SoC}. Security specification
mining focuses on developing the security goals of a design and has been
developed for processors~\cite{zhang2017identifyingshort,deutschbein2018mining,
  deutschbein2020CISC}.
However, many important vulnerabilities 
violate security goals related to how information flows, goals that are not expressible as the trace properties
that specification miners discover. 

The insight that led to this research is that mining the traces produced by an
IFT-instrumented design will generate trace properties that correspond to
information flow properties over the original, uninstrumented design. The
information flow tracking logic transforms the information-flow
properties from the space of hyperproperties~\cite{Clarkson2010} -- where
trace-based mining does not apply -- to the space of trace properties, where trace-based mining can apply.

A naive application of trace-based mining to an IFT-instrumented design quickly
runs into issues of complexity: the instrumented designs are large and 
overwhelm the miner. Additionally, the miner will discover properties over
tracking signals and original design signals that are meaningless and cannot be transformed back to the
space of information flow properties in the original design. To handle these issues we separate the process of identifying
source--sink flow pairs
in the design from the process of mining for the conditions that govern those
flows. The first can be done by leveraging existing information flow tracking tools
and the second makes use of existing trace miners. The key to making the
approach work is to synchronize the two parts using clock-cycle time.

The methodology we present here can inform an automated analysis of a
hardware design by identifying flow relations between all design
elements, 
including flow conditions and multi-source and multi-sink cases. The
methodology requires no input from the designer beyond the design and
testbench. 



To evaluate our methodology, we developed
\tool, a fully automatic security specification miner for information
flow properties. \tool uses information flow tracking (IFT)
technology from Tortuga Logic's Radix-S simulation based security
verification engine~\cite{Radix} and is implemented on top of Daikon~\cite{Ernst2007}, 
a popular invariant miner.

To our knowledge, \tool represents the 
first specification miner capable of extracting information flow security properties 
from hardware designs. 
Our results demonstrate:
\begin{itemize}
\item \tool characterizes the flow relations between all elements of a design.
\item \tool identifies important information flow security properties of a
  design without guidance from the designer.
\item \tool can be used to find undesirable flows of information in the design.
\item \tool is applicable to SoCs and CPUs.
\end{itemize} 

To measure our methodology and the usefulness of \tool's mined
specification, we evaluated \tool over an access control module,
a multi-controller and multi-peripheral system with a known security policy, and
a RISC-V design. We evaluated the output of \tool versus expected
information flow policies of the design and found information flow specifications
that, if followed, protect designs from known and potential future attack 
patterns.

 
\section{Properties}
\label{sec:props}

Isadora generates two styles of information flow properties: no-flow properties,
in which there is no flow of information between two design elements; and
conditional-flow properties, in which there exists some flow of information between
two design elements, but only when the design is in a certain state. Isadora can
also generate unconditional-flow properties, but these tend to be less
interesting for purposes of security validation.

\subsection{Tracking Information Flow}

 IFT can precisely measure all digital information flows in the
 underlying hardware, including, for example, implicit flows through
 hardware-specific timing channels.
  Isadora uses IFT at the
 register transfer level~\cite{Armaiti2017} to track data flow between registers
 rather than considering individual bits, with `registers' in this context
 referring to the Verilog notion of a register.
 Isadora may additionally be configured to consider Verilog wires,
 though doing so provided no observable improvements to generated specifications
 and considerably increased trace generation costs. The Isadora methodology
 can be applied to individual bits, as the underlying information
 flow tracking used within Isadora does consider individual bits.
 However, bit level analysis would result
  in extraordinarily high trace generation costs for even small designs.
 
Tracking proceeds as follows: for each signal
 $\mathtt{s}$ in
 the design, a new tracking
 signal $\mathtt{s}^T$ is added along with the logic needed to track how
 information propagates through a design. Once the tracking signals and
 tracking logic are added to the design, one or more signals may be set as the
 \emph{information source} by initializing their associated
 tracking signals to a nonzero value. All other tracking signals are initialized
 to zero. As the design executes, and information from a source signal
 propagates to a second signal, that second signal's tracking signal is updated
 from zero to nonzero.
 
 \subsection{Information Flow Restrictions}


 Using information flow tracking, we can express the property that information
 from register $\mathtt{r}_1$ should never flow to
 register $\mathtt{r}_2$ as a trace property: if $\mathtt{r}_1$ is the only signal whose tracking
 signal $\mathtt{r}_1^T$ is initialized to nonzero, then for all possible executions of the
 design, $\mathtt{r}_2$'s tracking signal $\mathtt{r}_2^T$ should remain at zero:
 \begin{align*}
&   (\forall \mathtt{r}_i, ~\mathtt{r}_i^T\neq 0 \leftrightarrow i = 1)
   \rightarrow \mathbb{G}(\mathtt{r}_2^T = 0)
 \end{align*}

This style of no-flow property can be useful for ensuring unprivileged users cannot
influence sensitive state or for ensuring that sensitive information cannot leak
through, for example, debug ports. 
However, it cannot capture conditional properties, for example that register
updates are allowed only under certain power states.

 \subsection{Information Flow Conditions}






 Using information flow tracking, we can express the property that information
 from a register $\mathtt{r}_1$ may flow to another
 register $\mathtt{r}_2$ under some condition $P$: if $\mathtt{r}_1$ is the only signal whose tracking
 signal $\mathtt{r}_1^T$ is initialized to nonzero then for all possible executions of the
 design, $\mathtt{r}_2$'s tracking signal $\mathtt{r}_2^T$ will only become non-zero
 if some predicate $P$ holds:
 \begin{align*}
&   
&  (\forall \mathtt{r}_i, ~\mathtt{r}_i^T\neq 0 \leftrightarrow i = 1)
   \rightarrow \mathbb{G}(\lnot P \rightarrow (\mathtt{r}_2^T = 0 \rightarrow \mathbb{X}(\mathtt{r}_2^T = 0)))
 \end{align*}

This style of a conditional flow property can be used to express, for example, that register updates are
allowable only under certain power states, or that memory accesses are allowable
only when specific access control checks have succeeded.


%

 \floatstyle{plain}
 \restylefloat{figure}
 \subsection{Grammar of Properties}


In order to produce properties that use only the signals in the original
design, without including the tracking signals, we need an operator that
expresses some notion of information flow. Both no-flow and conditional flow
properties can be expressed using a no-flow operator, for which we use the notation =/=>.
 The grammar of Isadora properties is as follows.



\begin{align*}
  \phi \doteq&~ \mathtt{r_1}\texttt{=/=> } \mathtt{r_2} ~|~ e \rightarrow \mathtt{r_1}
  \texttt{ =/=> } \mathtt{r_2} \\
  e \doteq&~ b \wedge e ~|~ b \\
  b \doteq&~ \mathtt{r} \in \{x, y, z\} ~|~
  \mathtt{r}_1 = \mathtt{r}_2~|~
  \mathtt{r}_1 \neq \mathtt{r}_2~|~
  \mathtt{r} = \mathrm{prev}(\mathtt{r})
  \end{align*}

The property $\mathtt{r_1}$ =/=> $\mathtt{r_2}$ states that no information flows
from $\mathtt{r_1}$ to  $\mathtt{r_2}$. The property $e \rightarrow \mathtt{r_1}
  \texttt{ =/=> } \mathtt{r_2}$ states that information may flow from
  $\mathtt{r_1}$ to  $\mathtt{r_2}$ only when $\neg e$. The symbol $\mathtt{r}$ is a register in the design, $\mathtt{r} \in \{x,
    y, z\}$ means that $\mathtt{r}$ may take on any one of the values in a set of cardinality
    less than or equal to three,
    and $\mathrm{prev}(\mathtt{r})$ refers to the value of $\mathtt{r}$ in the
    previous clock cycle.
    

\section{Methodology}

\begin{centering}
 \begin{figure}
 \includegraphics[width=\columnwidth]{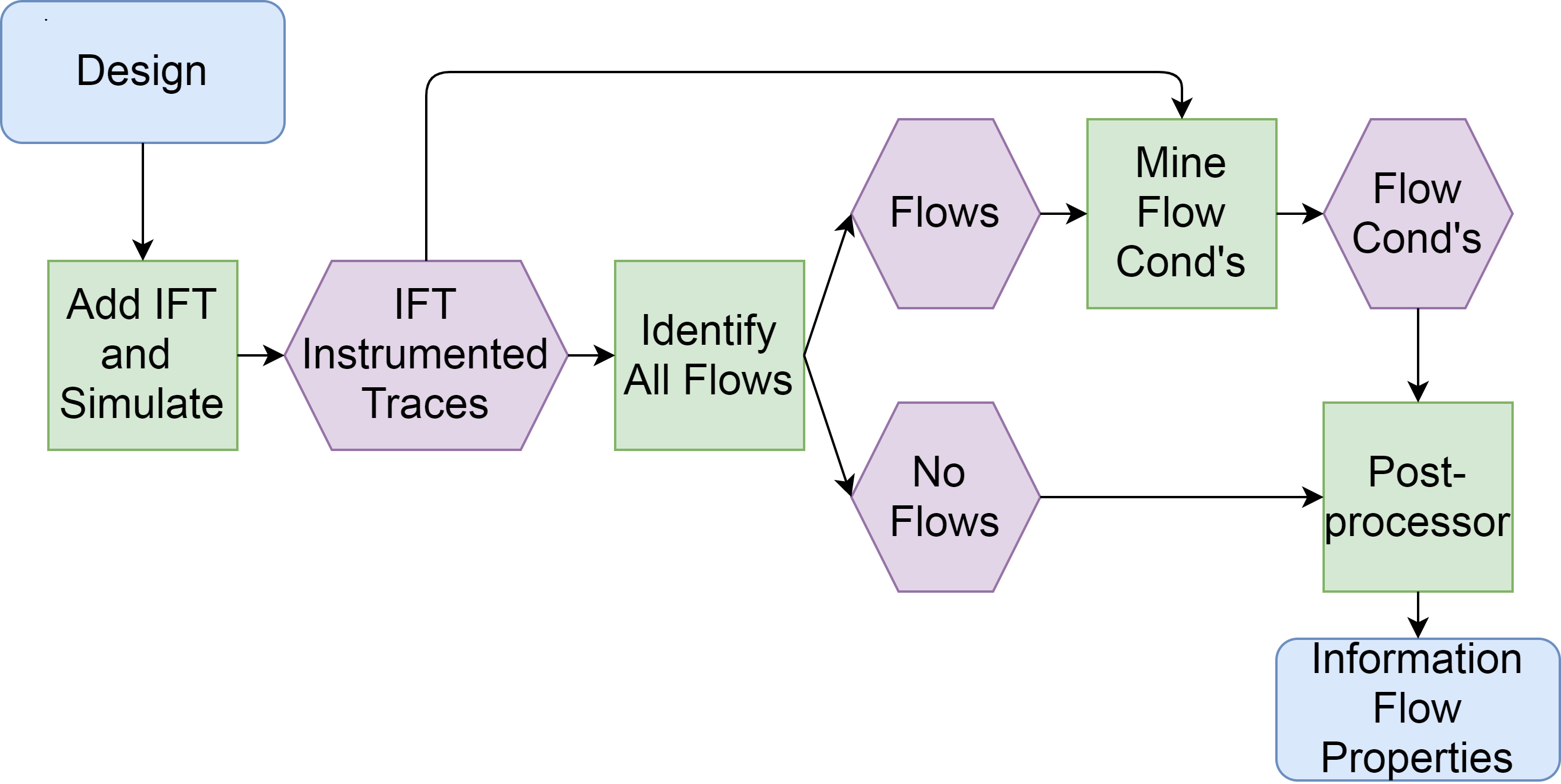}
 \caption{An overview of the Isadora workflow}
 \label{fig:overview}
 \end{figure}
 \end{centering}

Isadora analyzes a design in four phases: generating traces, identifying flows, mining for
flow conditions, and postprocessing.  An overview of
  the workflow is presented in Figure~\ref{fig:overview}. 
  
First, Isadora
instruments the design with IFT logic and runs the instrumented design in simulation using
the user-provided set of testbenches.
The result is a trace set 
that specifies the value
of every design signal and every tracking signal at each clock cycle during simulation.

Next, Isadora studies the trace set to find every flow that occurred during the simulation of the
design.
This set of flows is complete: if a flow occurred between any two signals,
it will be included in this set. At the end of this phase, Isadora also
produces the complete set of information flow restrictions: pairs of signals between which
no information flow occurs.

Then, Isadora uses an inference engine (Daikon~\cite{Ernst2007}) to infer,
for every flow that occurred, the predicates that specify the conditions under
which the flow occurred.

The final phase removes redundant and irrelevant predicates from the set
and logically combines the predicates with the information flows to produce the
conditional flow properties. These, along with the no-flow properties from the
second phase,
form the information flow specification produced by Isadora.

\subsection{Generating Traces with Information Flow Tracking}

To generate a trace set, the design is instrumented with IFT logic and then executed in
simulation with a testbench or sequence of testbenches providing input values to the design. 
Let $\tau_\mathtt{src} = \langle \sigma_0, \sigma_1, \ldots, \sigma_n \rangle$ be the
trace of a design instrumented to track how information flows from one signal,
$\mathtt{src}$, during execution of a testbench. The state $\sigma_i$ of
the design at time $i$ is defined by a list of triples describing the
current value of every design signal and corresponding tracking signal in the instrumented design:
\begin{align*}
  \sigma_i = &
[(\mathtt{s}_1,v_1, v^t_1), (\mathtt{s}_2,v_2, v^t_2), \ldots,
    (\mathtt{s}_m,v_m, v^t_m)]_i.
\end{align*}

In order to distinguish the source of a tainted sink signal, each input signal
must have a separate taint label. However, tracking multiple labels is
expensive~\citep{hu2014glift}. 
Therefore, Isadora takes a compositional approach. 
For each source signal, IFT instrumentation is configured to track the flow of information from only a single input signal of the
design, the $\mathtt{src}$ signal. This process is applied to each signal
in a design. 
The end result is a set of traces for design $\mathrm{D}$
and testbench $\mathrm{T}$: $\mathcal{T}_\mathrm{DT} =
\{\tau_\mathtt{src}, \tau_\mathtt{src'}, \tau_\mathtt{src''}, \ldots \}$. Each
trace in this set describes how information can flow from a single input signal to the rest
of the signals in the design. Taken together, this set of traces describes how
information flows through the design during execution of the testbench
$\mathrm{T}$.


\subsection{Identifying All Flows}
\label{sec:propertymining}
In the second phase, the set of traces are analyzed to identify:
\begin{enumerate}
\item every
pair of signals between which a flow occurs, and 
\item the
times within the trace at which each flow occurs.
\end{enumerate}
 Each trace
$\tau_\mathtt{src}$ is searched to find every state in which a
tracking signal goes from being set to 0 to being set to 1. In
other words, every signal--values triple $(\mathtt{s}, v, v^t)$ that is of the form
$(\mathtt{s}, v, 0)$ in state $\sigma_{i-1}$ and $(\mathtt{s}, v, 1)$ in state $\sigma_i$ is
found and the time $i$ is noted. This is stored as the tuple $(\mathtt{src}, \mathtt{s}, \{i_0, i_1, \ldots\})$, which indicates that information from
$\mathtt{src}$ reached signal $\mathtt{s}$ at all times $i \in \{i_0, i_1, \ldots\}$. 
We call this the
\emph{time-of-flow tuple}. 
There can be multiple times-of-flow within a single
trace because the tracking value of signals may be reset to zero by design events such as resets.

Once all traces have been analyzed, the collected time-of-flow tuples 
$(\mathtt{src}, \mathtt{s}, \{i_0, i_1, \ldots\})$ are organized by time. For any
given set of times $\{i_0, i_1, \ldots\}$ there may be multiple discovered flows. 
For all traces
$\tau_\mathtt{src}$ generated by a single testbench, the timing of flows from one source $\mathtt{src}$ can be
compared to the timing of flows from a second source $\mathtt{src'}$; the value
$i$ will refer to the same point in the testbench.
At the end of this phase, the tool produces two outputs. The first is a
list of the unique sets of times present within time-of-flow tuples
and all the corresponding register pairs for which flow is discovered
at precisely the times in the set:
\begin{align*}
S_\mathrm{flows} =  [\langle\{i_0, i_1, \ldots\}:
&\{(\mathtt{src_1}, \mathtt{s_1}), (\mathtt{src_2}, \mathtt{s_2}), 
\ldots\}\rangle;
\label{eq:flowset}\\
  \langle\{i_0', i_1', \ldots\}:
  &\{(\mathtt{src_1}', \mathtt{s_1}'), (\mathtt{src_2}', \mathtt{s_2}'), 
  \ldots\}\rangle; \ldots ]\nonumber.
\end{align*}
The same $\mathtt{src}$ may flow to many sinks 
$\mathtt{s} \in \{\mathtt{s_1}, \mathtt{s_2}, \ldots\}$ 
at the same times 
$i \in  \{i_0, i_1,  \ldots\}$,
and the same sink $\mathtt{s}$ may receive information from multiple sources 
$\mathtt{src} \in \{\mathtt{src_1}, \mathtt{src_2}, \ldots\}$ 
at the same times 
$i \in  \{i_0, i_1,  \ldots\}$.

The second output from this phase is a list of source-sink pairs between which
information never flows: 
\begin{equation*}
  S_\mathrm{no-flow} = \{(\mathtt{src}, \mathtt{s}), (\mathtt{src'}, \mathtt{s}'), \ldots\}. \label{eq:noflowset}
\end{equation*}

The pairs in this set comprise the noninterference properties of the design, and can be
specified using the no-flow operator, 
for example $\mathtt{src}$ \texttt{=/=>} $\mathtt{s}$

\subsection{Mining for Flow Conditions}
\label{sec:miningforflowconditions}
In the third phase, Isadora finds the conditions under which
a particular flow will occur. For example, if every time $\mathtt{src}$ flows to
$\mathtt{s}$, the register $\mathtt{r}$ has the value $x$, Isadora infers
the conditional information flow property:
 \begin{equation*}
\neg(\mathtt{r} = x) \rightarrow \mathtt{src} \texttt{ =/=> } 
 \mathtt{s} 
 \end{equation*}
Isadora uses the technique of dynamic invariant detection~\cite{Ernst2007} 
on traces to infer design behavior using
pre-defined patterns. 
In order to isolate the conditions for information flow between
two registers, Isadora uses $S_\mathrm{flows}$ to find all the
trace times $i$ at which information flows from $\mathtt{src}$ to $\mathtt{s}$ during
execution of the testbench. The corresponding trace(es) are then decomposed to
produce a set of trace slices that are two clock cycles in length, 
one for each time $i$. Consider
 time-of-flow tuple $(\mathtt{src}, \mathtt{s}, [i, j, k, \ldots])$, 
which as a notational convenience here uses distinct letters to denote time
points rather than subscripts for clarity in the
following expression. Given this tuple, Isadora
will produce the trace slices $\langle \sigma_{i-1}, \sigma_i \rangle, \langle
\sigma_{j-1}, \sigma_{j} \rangle, \langle \sigma_{k-1}, \sigma_{k}
\rangle$. These trace slices include only the signals of the original design,
all tracking logic and shadow signals are pruned.
Using trace slices, or trace windows of length two, allows 
dynamic invariant detection to generation predicates specifying
design state both immediately prior to and concurrent with the
occurence of some flow. Predicates match one of the four
patterns for expressions given in the grammar of 
Isadora properties in Section~\ref{sec:props}.

\subsection{Postprocessing}
\label{sec:isapost}

Finally, Isadora performs additional analysis to find invariants that may hold over the
entire trace set by running the miner on the unsliced trace. Isadora eliminates any predicate that is also found to be a trace-set
invariant. One such trivial example is the invariant $\mathtt{clk} = \{0,
1\}$. 

The final output properties from postprocessing are the conditional flow
properties. To ease readability, Isadora can express the conditional flow
properties as multi-source to multi-sink flows, where all flows within the same property 
occur at the same time and under
the same conditions. This produces comparatively few properties, which in practice
were approximately as many as the number of unique source signals, and avoids
redundant information.
The conditional flow properties and the no-flow properties discovered in phase 2
(Section~\ref{sec:propertymining}) make up the set of information flow properties produced by Isadora.
Two examples of postprocessed properties are shown in Appendix~\ref{sec:samples}.


\section{Implementation}

Isadora uses the Tortuga Radix-S simulation-based
security verification technology~\cite{Radix} to generate IFT logic for a
hardware design, the Questa Advanced Simulator~\cite{questasim} to simulate the instrumented
design and generate traces, and the Daikon~\cite{Ernst2007}
invariant miner to find flow conditions. A Python
script manages the complete workflow and implements flow analysis and 
postprocessing. 


Traces are generated for all signals within a
design. An automated utility identifies every signal within a design and
configures Tortuga Radix-S to build the IFT logic separately for each of these
registers. We run Tortuga in 
exploration mode, which omits cone of influence analysis, and track flows
to all design state using the \texttt{\$all\_outputs} variable. The resulting instrumented designs are simulated in QuestaSim over a
testbench (see Evaluation, Sec.~\ref{sec:evaluation}) to produce a trace of
execution.

Phase two is implemented as a Python tool that reads in the traces generated by
QuestaSim and produces the set of no-flow properties and the 
time-of-flow tuples. This phase combines the bit-level taint
tracking by Radix-S into signal-level tracking. 
Each $n$-bit signal in the original design is then
tracked by a 1-bit shadow signal, which will be set to 1 at the first point in
the trace that any of the component
$n$ shadow bits where set.

The mining phase is built on top of the Daikon invariant generation 
tool~\cite{Ernst2007}, which was developed for use with software programs. 
Daikon generates invariants over state variables for each point in a program. We
built a Daikon front-end in Python (411 LoC, including comments) 
that converts the trace data to be Daikon readable, 
treating the state of the design at each clock cycle as a point in a
program. The front-end also removes any unused or redundant signals and 
outputs relevant two-clock-cycle slices as described
 in Sec.~\ref{sec:miningforflowconditions}.

\section {Evaluation}
\label{sec:evaluation}

\newcommand{\refset}{designer--provided, assertion--based,
information--flow security specification}

We assess the following questions to evaluate Isadora:
\begin{enumerate}
    \item
Can Isadora independently mine security properties manually developed
by hardware designers?
    \item
Can Isadora automatically generate properties describing Common Weakness Enumerations (CWEs)~\cite{CWE} over a
design?
    \item
Does Isadora scale well for larger designs, such as CPUs or SoCs?
\end{enumerate}

\subsection{Designs}
We assessed Isadora on two designs, the Access Control Wrapper (ACW) proposed
within the AKER framework~\citep{Restuccia2021AKER} and the PicoRV32  RISC-V CPU.
An ACW wraps an AXI controller and enforces on it a \emph{local access control
policy}, which is setup and maintained by a trusted entity (e.g., a Hardware
Root of Trust or a trusted processor).
The ACW checks the validity of read and write requests issued by the wrapped AXI controller and rejects those that violate the configuration of the local access control policy.

\begin{figure}
\centering
\includegraphics[width=\columnwidth]{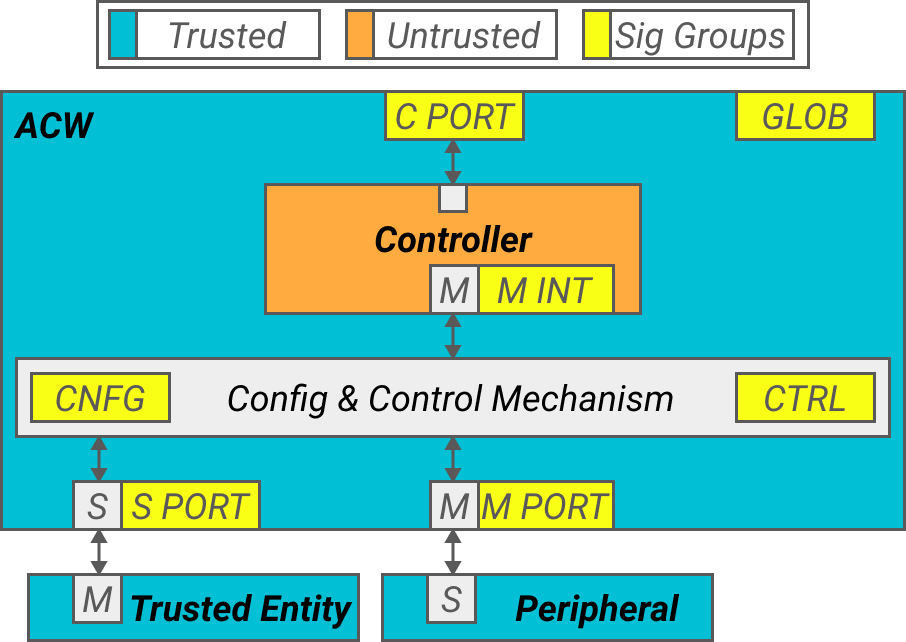}
\caption{Block diagram of the Single ACW design, with labeled signal groups.}
\label{fig:blocks}
\end{figure}

\begin{figure}
\centering
\includegraphics[width=\columnwidth]{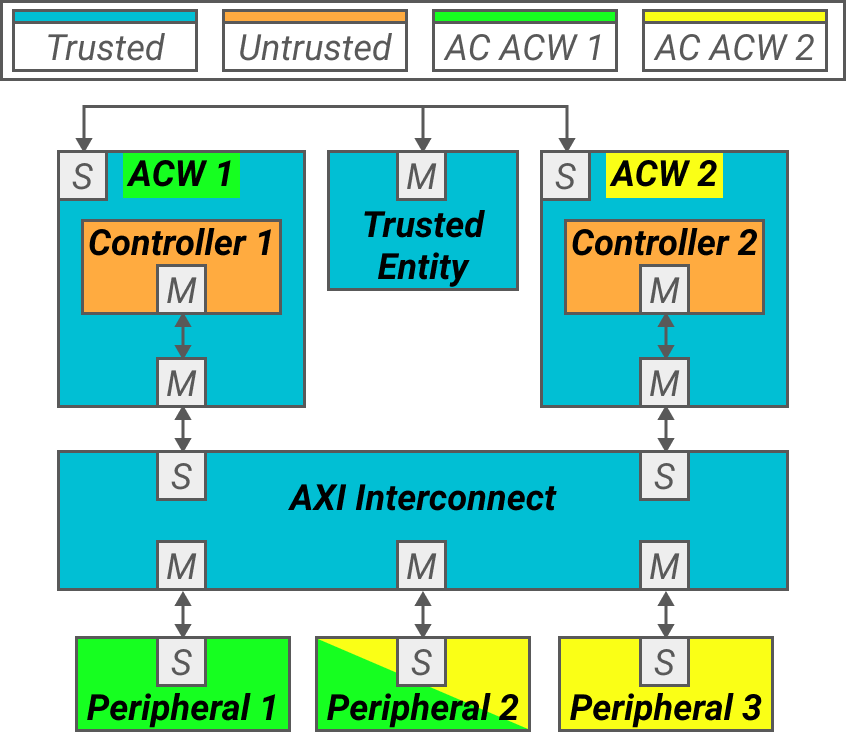}
\caption{Block diagram of the Multi ACW design}
\label{fig:multiACW}
\end{figure}

We used the AKER framework in two configurations: first
implementing a single-controller AKER-based access control system; second,
implementing a system with two traffic generators, each wrapped by an ACW,
connected to three AXI peripherals though an AXI interconnect. This setup simulates
the use of the ACWs in an SoC environment.
In both cases, the input signals of the ACWs are dictated by the testbench, which initializes them with the access control policies and acts as the trusted entity.
We refer to these two designs as the \emph{``Single ACW''}
and \emph{``Multi ACW''} cases. They are shown in Figures~\ref{fig:blocks} and~\ref{fig:multiACW}, respectively.

PicoRV32 is a CPU core that implements the RISC-V RV32IMC Instruction Set, an open standard instruction set architecture
based on established reduced instruction set computer principles.

The secure operation of the ACW and AKER-based access control systems has been
verified through a property-based security validation process by the designers.
We study the AKER framework to evaluate how Isadora's properties compare to a
manually developed security specification. We use the PicoRV32 to evaluate how well Isadora automatically generates properties describing CWEs and to evaluate how well Isadora scales on a CPU design.

%


\subsection{Time Cost}

Isadora ran on a system with an
Intel Core i5-6600k (3.5GHz) processor with 8 GB of RAM. Traces were generated
on a Intel Xeon CPU E5-2640 v3 @ 2.60GHz server. Trace generation
dominated time costs, and scaled slightly worse than linear with number of
unique signals in a design.
Trace generation was suitable for parallelization though parallelization
was not considered in the evaluation.

The design sizes are given in Table~\ref{tab:sizetable}.
For the Single ACW, trace generation took
9h33m.
For the Multi ACW, trace generation exceeded
24 hours so we consider a reduced trace, which tracks sources
for one of the ACWs, though all signals are included
as sinks or in conditions.
The reduced trace was generated in 6h48m.
For PicoRV32, trace generation took
8h35m.

\begin{table*}
\centering
\begin{tabular}{ l r r r r r r r r }
\toprule
 Design & Unique& Unique & LoC & Trace &
Trace & Daikon  & Isadora  & Miner Time\\
&  Signals & Sources &  & Cycles & GBs  & Traces& Properties & In Minutes \\
\midrule
Single ACW & 229 & 229 & 1940 & 598 & .7 & 252 & 303 & 29:51 \\
Multi ACW & 984 & 85 & 4447 & 848 & 4.3 & 378 & 160 & 8:31 \\
PicoRV32 & 181 & 181 & 3140 & 1099 & .6 & 955 & 153 & 15:09 \\
\bottomrule
  \end{tabular}
\caption{Various size measures of studied designs}
\label{tab:sizetable}
\end{table*}

\subsubsection{Theoretical Gains to Parallelization}

When parallelizing all trace generation and all case mining, Isadora could
theoretically evaluate the Single ACW case fully in less than five minutes.
Parallelizing the first phase requires a Radix-S and QuestaSim
instance for each source register,
and each trace is generated in approximately 100 seconds.
Further, the trace generation time is dominated by write-to-disk, and
performance engineering techniques could likely reduce it significantly,
such as by changing trace encoding or piping directly to later phases.
Parallelizing the second phase
requires only a Python instance for each source register,
and takes between 1 and 2 seconds per trace.
Parallelizing the third phase requires a Daikon
instance for each flow case, usually roughly the same number as unique sources,
and takes between 10 and 30 seconds per flow case.
The final phase, postprocessing, is also suitable for parallelization.
Maximally parallelized, this gives a
design-to-specification time of under four minutes for the single ACW and for
similarly sized designs, including PicoRV32.

\subsection{Designer Specified Security Properties}
\label{sec:refsetprops}

\begin{table*}
\centering
\begin{tabular}{c|c|c|c|c|c|c}
\hline
   Source & Sink & Invariant & Provided & Result & Isadora  & CWEs  \\
    &  &  &  Assert's & & Properties &   \\
    \hline
M PORT\rule{0pt}{2.2ex} & M INT & GLOB  & 19 & \ding{51} & 2, 40,   & 1258, 1266, 1270,   \\
\cline{0-1}
\cline{4-5}
M INT\rule{0pt}{2.2ex} & M PORT &   & 19 & \ding{51} &  43, 53, & 1271, 1272, 1280\\
\cline{0-4}
\cline{7-7}
M PORT\rule{0pt}{2.2ex} & M INT & C PORT  & 19 & \ding{51} &  54, 204, & 1258, 1270, \\
\cline{0-1}
\cline{4-5}
M INT\rule{0pt}{2.2ex} & M PORT &  & 19 & \ding{51} & 214 & 1272, 1280\\
\hline
S PORT\rule{0pt}{2.2ex} & CNFG & - & 4 &  \ding{55} & 2, 6 & 1269, 1272, 1280\\
\hline
\end{tabular}
\caption{Isadora performance versus manual specification, on the Single ACW}
\label{tab:singleACW}
\end{table*}

For the Single ACW we compared Isadora's output against
security assertions developed by the AKER~\cite{Restuccia2021AKER} designers using
the Common Weakness Enumerations (CWE) database~\cite{CWE} as a guide.
These assertions, the CWEs described, and the results of Isadora on the
Single ACW
are shown in Table ~\ref{tab:singleACW}.
For each assertion
Isadora mined either a property containing the assertion
or found both a violation and the violating conditions for each assertion.
We reported
the observed violations to the designers who determined that the design
remained secure but a conditional flow had been incorrectly specified as
always illegal. Isadora also found the
conditions for legality.

Only 9 Isadora properties, out of 303 total Isadora properties generated,
were required to cover the designer-provided assertions,
including conditions specifying violations.
The Isadora output properties may contain many source or sink signals that flow
concurrently and their corresponding
conditions, whereas the designers' assertions each considered a single source and sink.
For example, on the ACW nine distinct read channel registers always flow to a
corresponding read channel output wire at the same time,
so Isadora outputs a single property for this design state.
This state included the reset signal and a
configuration signal both set to non-zero values, which were captured
as flow conditions, demonstrating
correct design implementation. This single
Isadora property captured 18 low level assertions related to multiple CWEs.

\subsubsection{Case Study: Unintended Proxy}

In the Multi ACW case, we studied
CWE 411: Unintended Proxy or Intermediary
(`Confused Deputy'). The system contained two controllers (\textit{C}), with two
access control modules (\textit{ACW}), a trusted entity that configured each ACW
(\textit{T}),
and three peripherals (\textit{P}). The ACWs each implemented an
access control (\textit{AC}) policy shown in Figure~\ref{fig:multiACW}
and given as:
\begin{align*}
\mathrm{AC}_1 \; \mathrm{of}\; \mathrm{ACW}_1:&\quad  R = \{P_1, P_2\},\: W = \{P_1\}\\
\mathrm{AC}_2 \; \mathrm{of}\; \mathrm{ACW}_2:&\quad  R = \{P_3\},\: W = \{P_2, P_3\}
\end{align*}
Isadora discovered legal flows from the $ACW_2$ write data to $P_3$ read and write data,
and  $P_2$ read data. Isadora also finds an illegal flow
to $P_1$ write data. The $ACW_2$-to-$P_1$ illegal flow has a flow condition specifying a prior flow from the
relevant signals within $ACW_2$ to $ACW_1$. While not constituting a precise path
constraint, this captures an access control violation and suggests the
confused deputy scenario because the flow profile from $ACW_2$ is consistent
with this path.

\subsection{Automatic Property Generation}
\label{sec:isadoraeval2}

For the two designs with full trace sets, the Single ACW and PicoRV32, Isadora
generates a specification describing all information flows and their conditions
with hundreds
of properties. To assess whether these properties are security properties, for
each design we randomly selected 10 of the 303 or 153 total properties
(using Python random.randint) and assessed their relevance to security.

We use CWEs as a metric to evaluate the security relevance of Isadora
output properties.
To do so, for each design, we first determine which
CWEs apply to the design. For both the ACW and PicoRV32, we
used the Radix Coverage for Hardware Common Weakness Enumeration (CWE)
Guide~\cite{Radix} to provide a list of CWEs that
specifically apply to hardware. We considered each documented CWE for
both designs. CWEs, while design agnostic, may refer to design features not
 present in the Single ACW or PicoRV32 or may not refer to
information flows. High level descriptions in multiple CWEs may
 correspond to the same low level behavior for a design and
 we consider these CWEs together.

Information flow hardware CWEs describe source signals, sink signals,
and possibly conditions.
CWEs provide high level descriptions, but Isadora targets
an RTL definition.
To apply these high level descriptions to RTL, we first group signals for a design
by
inspecting Verilog files and, if available, designer
notes. With the groups established, we
label every property by which group-to-group flows they contain.
We also determine which source--sink flows could be
described in CWEs, which often correspond to, or even match exactly, a signal group.
We use these groups to find CWE-relevant,
low-level signals as sources, sinks, and conditions in an Isadora property.
We also use these groups to characterize the relative frequency of
conditional flows between different groups, which we present as heatmaps in the
following subsections.

\subsubsection{ACW Conditional Information Flow}
\label{sec:acmcweeval}

Over the ACW we assess fourteen CWEs which we map to five plain-language
descriptions of the design features, as shown in
Table~\ref{tab:ACMCWEgroups}.

\begin{table}
\centering
  \begin{tabular}{ l l   }
    \midrule
CWE(s) & Description  \\ \midrule
1220 & Read/write channel separation  \\
1221-1259-1271 & Correct initialization, reset, defaults \\
1258-1266-1270-1272 & Access controls use operating modes  \\
1274-1283 & Anomaly registers log transactions \\
1280 & Control checks precede access \\
1267-1269-1282 & Configuration/user port separation  \\
    \midrule
  \end{tabular}
\caption{
The 14 CWEs considered for ACW}
\label{tab:ACMCWEgroups}
\end{table}

For the ACW signal groups, all registers were helpfully placed into
groups by the designer and labeled within the design. The design contained
seven distinct labeled groups:
\begin{itemize}
\item `GLOB' - Global ports
\item `S PORT' - AXI secondary (S) interface ports of the ACW
\item `C PORT' - Connections to non-AXI ports of the controller
\item `M PORT' - AXI main (M) interface ports of the ACW
\item `CNFG' - Configuration signals
\item `M INT' - AXI M interface ports of the controller
\item `CTRL' - Control logic signals
\end{itemize}

GLOB signals are clock, reset, and interrupt lines.
S PORT represents the signals that the trusted entity \textit{T} uses to configure the ACW.
C PORT represents the signals which are used to configure the controller \textit{C} to generate traffic for testing.
M PORT carries traffic between the peripheral \textit{P} and the ACW’s control mechanism.
CNFG represents the design elements which manage and store the configuration of the ACW.
M INT carries the traffic between the ACW’s control mechanism and the controller.
If it is legal according to the ACW’s configuration, the control mechanism will send M INT traffic to M PORT and vice versa.
CTRL represents the design elements of the aforementioned control mechanism.

\begin{figure}
\centering
\includegraphics[width=\columnwidth]{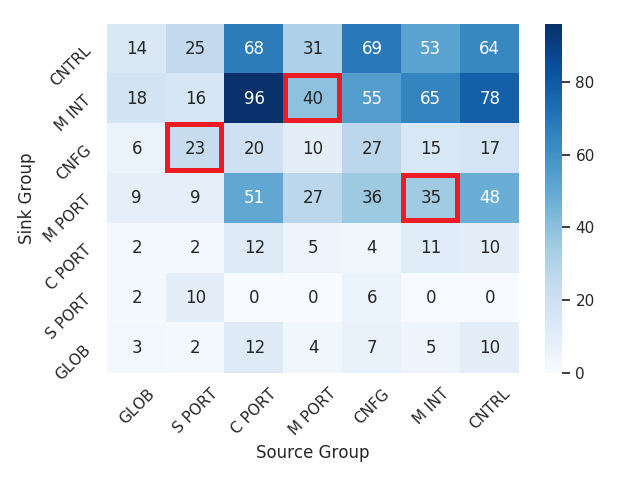}
\caption{Group-to-group conditional flow heatmap for the Single ACW.}
\label{fig:heatmap}
\end{figure}

First consider the heatmap view of the Single ACW in Figure~\ref{fig:heatmap}.
In this view, all of the designer-provided
assertions fall into just 3 of the 49 categories; these are outlined
in red. Further, all of the violations
were found with S PORT to CNFG flows, while
all satisfied assertions were flows between M INT and M PORT.
Another interesting result visible in the heatmap is the
infrequent flows into S PORT, which is used by the trusted entity to
program the ACW. Most of the design features should not be able to reprogram
the access control policy, so finding no flows along these cases provides
a visual representation of secure design implementation with respect to these
features.

For the ACW, all ten sampled properties encode CWE-defined behavior to
prevent common weaknesses, as shown in Table~\ref{tab:CWEACM}. In this table,
the columns labeled by a CWE number and a `+' refer to all the CWEs given
in a row of Table~\ref{tab:ACMCWEgroups}.
Eight out of the ten properties provide separation between read and write channels, which constitutes the main
functionality of the ACW module. CWEs 1267, 1269, and 1282 are not found within the
conditional flow properties produced by Isadora as these are no-flow properties,
so they are not present within the samples drawn from the numbered, conditional flow
properties, but we were able to verify they are included in Isadora's set of
no-flow properties.

\begin{table*}
\centering
\begin{tabular}{rllllll}
\toprule
   \# & Description & 1220 & 1221+ & 1258+ & 1274+ & 1280 \\
\midrule
3 & Control check for first read request after reset &
 \checkmark & & \checkmark & & \checkmark \\
10 & Secure power-on &
 & \checkmark &  & &  \\
37 & Anomalies and memory control set after reset &
 \checkmark & & \checkmark &  \checkmark & \checkmark \\
96 & \textit{T} via S PORT configures ACW &
 \checkmark & & &  \checkmark & \checkmark \\
106 & Interrupts respect channel separation &
 \checkmark & & & & \\
154 & Base address not visible to \textit{P} during  reset &
  & & \checkmark & & \\
163 & Write transaction legality flows to \textit{P}  &
\checkmark  & &  & & \\
227 & Write channel anomaly register updates &
\checkmark  & &  & \checkmark & \\
239 & Write validity respects channel separation, reset &
\checkmark  & &   \checkmark & & \\
252 & Read validity respects channel separation, reset &
\checkmark  & &   \checkmark & & \\
\bottomrule
\end{tabular}
\caption{Sampled Isadora properties on Single ACW}
\label{tab:CWEACM}
\end{table*}

\subsubsection{PicoRV32 Conditional Information Flow}

Over PicoRV32 we assess eighteen CWEs which we map to seven plain language
descriptions of the design features, as shown in
Table~\ref{tab:RISCVCWEgroups}.

\begin{table}
\centering
  \begin{tabular}{ l l  }
    \midrule
CWE(s) & Description  \\ \midrule
276-1221-1271 & Correct initialization, reset, defaults \\
440-1234-1280-1299 &  Memory accesses pass validity checks \\
1190 & Memory isolated before reset \\
1191-1243-1244... & Debug signals do not interfer with  \\
-1258-1295-1313 & ...any other signals \\
1245 & Hardware state machine correctness \\
1252-1254-1264 & Data and control separation \\
    \midrule
  \end{tabular}
\caption{
The 18 CWEs considered for PicoRV32}
\label{tab:RISCVCWEgroups}
\end{table}

PicoRV32 had no designer-specified signal groups so we used
comments in the code,
register names, and code inspection to group all signals. We use
lower case names to denote these groups were not defined by the designer.
\begin{itemize}
\item `out' - Output registers
\item `int' - Internal registers
\item `mem' - Memory interface
\item `ins' - Instruction registers
\item `dec' - Decoder
\item `dbg' - Debug signals and state
\item `msm' - Main state machine
\end{itemize}
The memory interface and the main state machine were indicated
by comments in the code. The instruction registers, the decoder, and debug all appeared under one disproportionately large section described
as the instruction decoder. Debug was grouped by name after
manual analysis found registers in this region prefixed with `dbg\_',
`q\_', or `cached\_' to interact with and only with one another.
Instruction registers prefixed
`instr\_' all operate similarly to each other
and differently than the remaining decoder signals, which were placed
in the main decoder group.
Internal signals were
the remaining unlabeled signals that appeared early within the design, such
as program and cycle counters and interrupt signals, and the output registers
were all signals declared as output registers.

\begin{figure}
 \centering
 \includegraphics[width=\columnwidth]{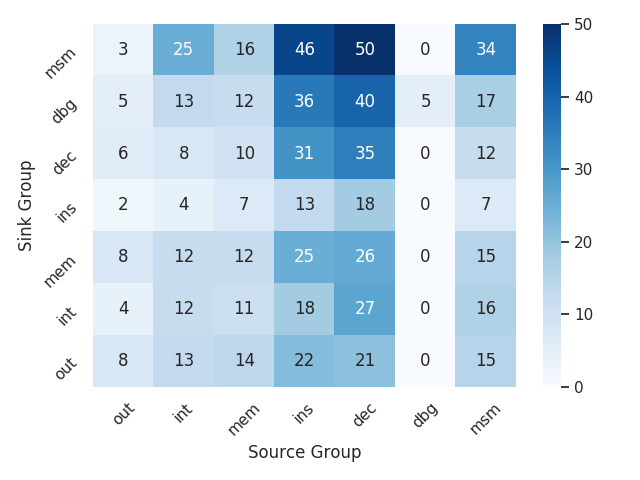}
 \caption{Group-to-group conditional flow heatmap for PicoRV32. }
 \label{fig:heatmapr5}
 \end{figure}

First consider the heatmap view of PicoRV32 in Figure~\ref{fig:heatmapr5}.
An interesting result visible in the heatmap is the flow isolation from debug
signals to the rest of the design. Many exploits, both known and anticipated,
target debug  information leakage, and the heatmap shows this entire class of
weakness is absent from the design.

For PicoRV32 we find eight of ten sampled properties encode CWE defined behavior to
prevent common weaknesses. We present these results in Table~\ref{tab:CWERISCV}.
The columns labeled by a CWE number and a `+' refer to all the CWEs given
in a row of Table~\ref{tab:RISCVCWEgroups}.
The remaining two Isadora properties were single source or single sink
properties representing a logical combination inside the decoder, and
captured only functional correctness.

\begin{table*}
\centering
\begin{tabular}{rlllllll}
\toprule
   \# & Description & 276+ & 440+ & 1190 & 1191+ & 1245 & 1252+ \\
\midrule
1 & No decoder leakage via debug &
 & &  & \checkmark & \\
16 & Instructions update  state machine &
 & \checkmark &  & & \checkmark \\
30 & Decoder updates state machine &
 & \checkmark \\
47 & No state machine leakage via debug &
 & & &  \checkmark & \\
52 & SLT updates state machine &
& & & &  \checkmark \\
66 & Handling of jump and load &
  & & \checkmark & \checkmark & & \checkmark \\
79 & Loads update state machine &
 &  &  & & \checkmark \\
113 & Decoder internal update &
  & &  & & \\
130 & Write validity respects reset &
 &  &  & & \checkmark \\
144 & Decoder internal update &
  & &  & & \\
\bottomrule
\end{tabular}
\caption{Sampled Isadora properties on PicoRV32}
\label{tab:CWERISCV}
\end{table*}

\section{Discussion}

In this section, we discuss the threats to validity for
properties produced using Isadora.

False positives may be introduced by
 insufficient trace coverage, by
limitations of information flow tracking, or by incorrectly classifying
functional properties
as security properties. Sampling output properties found a
10\% false positivity rate with respect to misclassification. This rate is discussed
in greater detail in Section~\ref{sec:functionalpropsisadora}.

With regard to false negatives,
they fall into two cases: known and unknown.
Isadora captured all known assertions for the Single ACW (Section~\ref{sec:refsetprops}). 
For Single ACW and PicoRV32 properties (Section~\ref{sec:isadoraeval2}), the sampled properties
partially addressed
all CWEs manually determined to be relevant, 
but no CWE was completely covered within the
sampled properties. 
A manual inspection 
is needed to rule out the possibility of false negatives with respect to CWEs.
Unknown false negatives can arise from limitations in trace
coverage or in logical specificity.

\subsection{Trace Reliance}

As a specification miner, Isadora relies on traces.
The second stage
relies on traces with sufficient case
coverage to drive information flow through all
channels present in the design.
The third stage relies on traces to infer flow predicates.
Over buggy hardware,
these predicates may form a specification describing buggy behavior.
Traces may not cover all cases that can be reached by a design or even
occur during normal design operation.

Traces may not precisely describe some design features.
For example, when considering property number 154 
on the Single ACW, one of the sampled properties,
Isadora found
predicates that ARLEN\_wire and AWLEN\_wire are both set to be exactly 8 for
any flow to occur. 
This property is shown in full in Appendix~\ref{sec:sampleoutput}.

The AxLEN\_wire registers set transaction burst size for reads and writes. For
transactions in write channels, the ARLEN\_wire value
should be irrelevant,
 and this clause within the broader property constitutes
a likely false positive.

The AWLEN\_wire is a different case. In a properly configured write channel
supporting transcations, this register would necessarily be non-zero, and for
wrapping bursts must be a power of two,
but manual inspection of the code provides no indication the value must be
precisely 8. 

While Isadora is testbench reliant,
testbench generation is an active
area of research, and is more fully explored in related works
such as Meng et al.~\cite{Meng2021Concolic}, which studies
concolic testing for RTL.

\subsection{Functional Properties}
\label{sec:functionalpropsisadora}

When using CWE-relevance as the metric, Isadora does include functional
properties in its output, as shown in Table~\ref{tab:CWERISCV}.
Sampling output properties found a
10\% false positivity rate with respect to misclassification for
the sampled properties from both designs, with 0 of 10 properties
found to be false positives over the Single ACW version of AKER, and
2 of 10 properties found to be false positives over the PicoRV32
RISC-V CPU.

We attribute finding functional properties solely on RISC-V primarily to differences in
design and testbench. 
The ACW studied was the target of validation efforts related to information flow,
and the testbench we used was developed as part of those efforts.
Further, as an access control module, by nature much of its functionality
was relevant to secure access control.

With RISC-V, a minimal test bench was used that was intended
only to run the design in an environment without access to the full
RISC-V toolchain (such as our simulation environment for instrumented
trace generation), and much of the design was devoted to behavior for
which CWEs did not apply, such
as logical updates during instruction decoding.
One example of an Isadora property classified as functional is shown in Appendix~\ref{sec:case144}.
%
%
%

\subsection{Measuring Interference}

Isadora assumes the correctness of the information flow tracking used
in trace generation. Information flow tracking is an active
area of research, and is more fully explored in related works
such as Ardeshiricham et al.~\cite{Armaiti2017}, which studies
IFT for RTL.

%

\subsection{Specification Logic}

Isadora does not define temporal properties beyond a single delay slot
incorporated in the trace slices of length two. However,
manual examination of output properties suggests information flow
patterns during initialization,
which is the first 4 cycles for AKER and first 80 for RISC-V,
are highly dissimilar to later flows.
During initialization, Isadora discovers
flow conditions referencing registers with unknown
states.
Isadora also finds
concurrent flows between elements for which no concurrent flows occur
after reset.
Because conditions are inferred from comingled trace slices from during
and after initialization, the output properties may be insufficiently
precise to capture secure behavior related to this boundary.

\section{Related Work}

\subsection{Properties of Hardware Designs}
Automatic extraction of security critical assertions from hardware
designs enables assertion based verification without first
manually defining properties~\cite{hu2016towards}. IODINE looks for possible
instances of known design patterns, such as one-hot encoding or mutual exclusion
between signals.~\cite{HangalDAC2005}. More recent papers use data mining of simulation
traces to extract more detailed assertions~\cite{ChangASPDAC2010,HertzTCAD2013}
or temporal properties~\cite{li2010scalable}. Recent work
has focused on mining temporal properties from execution
traces~\cite{li2010scalable,danese2017team,danese2015automaticdate,danese2015automatic}. 
A combination of static and dynamic analysis extracts word-level properties~\cite{liu2012word}.

The first security properties developed for hardware designs were manually
crafted~\cite{HicksASPLOS2015,IrvineHOST2011,MichaelThesis}. 
SCIFinder semi-automatically generates security-critical properties for a 
RISC processor
design~\cite{zhang2017identifyingshort} and Astarte generates security-critical
properties for x86~\cite{deutschbein2020evaluatingshort}. Recent hackathons have revealed the types of
properties needed to find exploitable bugs for RISC SoCs~\cite{dessouky2019hardfails}. 

\subsection{Mining Specifications for Software}
The seminal work in specification mining comes from the software
domain~\cite{ammons2002mining} in which execution traces are examined to infer
temporal specifications as regular expressions. Subsequent work used
both static and dynamic traces to filter candidate
specifications~\cite{weimer2005mining}. More recent work has tackled imperfect execution traces~\cite{yang2006perracotta},
and the complexity of the search
space~\cite{gabel2008javert,reger2013pattern,gabel2008symbolic}.
Daikon, which produces invariants rather than temporal properties, learns
properties that express desired semantics of a program~\cite{Ernst2007}.

In the software domain a number of papers
have developed security specific specification mining tools. These tools use
human specified rules~\cite{TanUSENIXSec2008}, observe instances of deviant
behavior~\cite{PerkinsSOSP2009, MinSOSP2015, Ernst2007}, or identify
instances of known bugs~\cite{YamaguchiWOOT2011}.

\section{Conclusion}

We presented and implemented a methodology for creating information flow specifications of hardware designs. 
By combining information flow tracking and specification mining, we are able to produce information flow properties of a design without prior knowledge of security agreements or specifications. 
We show our implementation, Isadora, characterizes the flow relations between
all elements of a design and identifies important information flow security
properties of an SoC and a CPU according to Common Weakness Enumerations.

\section{Acknowledgments}
We thank our reviewers for their insightful comments and suggestions. This material is based upon work supported by the National Science 
Foundation under Grants No. CNS-1816637 and 1718586, by the Semiconductor
Research Corporation, and by Intel. Any opinions, findings, conclusions, 
and recommendations expressed in this paper are solely those of the authors.

\bibliographystyle{ACM-Reference-Format}
\bibliography{references}


\begin{thebibliography}{45}


\ifx \showCODEN    \undefined \def \showCODEN     #1{\unskip}     \fi
\ifx \showDOI      \undefined \def \showDOI       #1{#1}\fi
\ifx \showISBNx    \undefined \def \showISBNx     #1{\unskip}     \fi
\ifx \showISBNxiii \undefined \def \showISBNxiii  #1{\unskip}     \fi
\ifx \showISSN     \undefined \def \showISSN      #1{\unskip}     \fi
\ifx \showLCCN     \undefined \def \showLCCN      #1{\unskip}     \fi
\ifx \shownote     \undefined \def \shownote      #1{#1}          \fi
\ifx \showarticletitle \undefined \def \showarticletitle #1{#1}   \fi
\ifx \showURL      \undefined \def \showURL       {\relax}        \fi
\providecommand\bibfield[2]{#2}
\providecommand\bibinfo[2]{#2}
\providecommand\natexlab[1]{#1}
\providecommand\showeprint[2][]{arXiv:#2}

\bibitem[\protect\citeauthoryear{MITRE}{CWE}{[n.d.]}]%
        {CWE}
 \bibinfo{year}{[n.d.]}\natexlab{}.
\newblock \bibinfo{title}{{ The Common Weakness Enumeration Official Webpage}}.
\newblock
\newblock
\urldef\tempurl%
\url{https://cwe.mitre.org/}
\showURL{%
\tempurl}


\bibitem[\protect\citeauthoryear{??}{que}{[n.d.]}]%
        {questasim}
 \bibinfo{year}{[n.d.]}\natexlab{}.
\newblock \bibinfo{title}{{Questa Advanced Simulator}}.
\newblock
\newblock
\urldef\tempurl%
\url{https://eda.sw.siemens.com/en-US/ic/questa/simulation/advanced-simulator/}
\showURL{%
\tempurl}


\bibitem[\protect\citeauthoryear{??}{Rad}{[n.d.]}]%
        {Radix}
 \bibinfo{year}{[n.d.]}\natexlab{}.
\newblock \bibinfo{title}{{Radix Coverage for Hardware Common Weakness
  Enumeration (CWE) Guide}}.
\newblock
\newblock
\newblock
\shownote{\url{https://tortugalogic.com/wp-content/uploads/2020/03/RadixCWEGuide_20210126.pdf}.}


\bibitem[\protect\citeauthoryear{Ammons, Bod\'{\i}k, and Larus}{Ammons
  et~al\mbox{.}}{2002}]%
        {ammons2002mining}
\bibfield{author}{\bibinfo{person}{Glenn Ammons}, \bibinfo{person}{Rastislav
  Bod\'{\i}k}, {and} \bibinfo{person}{James~R. Larus}.}
  \bibinfo{year}{2002}\natexlab{}.
\newblock \showarticletitle{Mining Specifications}. In
  \bibinfo{booktitle}{\emph{29th Symposium on Principles of Programming
  Languages (POPL)}}. \bibinfo{publisher}{ACM}, \bibinfo{pages}{4--16}.
\newblock
\showISBNx{1-58113-450-9}
\urldef\tempurl%
\url{https://doi.org/10.1145/503272.503275}
\showDOI{\tempurl}
\newblock
\shownote{\url{http://doi.acm.org/10.1145/503272.503275}.}


\bibitem[\protect\citeauthoryear{Ardeshiricham, Hu, and Kastner}{Ardeshiricham
  et~al\mbox{.}}{2017a}]%
        {armaiti2017clepsydra}
\bibfield{author}{\bibinfo{person}{Armaiti Ardeshiricham}, \bibinfo{person}{Wei
  Hu}, {and} \bibinfo{person}{Ryan Kastner}.} \bibinfo{year}{2017}\natexlab{a}.
\newblock \showarticletitle{Clepsydra: Modeling timing flows in hardware
  designs}. In \bibinfo{booktitle}{\emph{IEEE/ACM International Conference on
  Computer-Aided Design (ICCAD)}}. \bibinfo{pages}{147--154}.
\newblock
\urldef\tempurl%
\url{https://doi.org/10.1109/ICCAD.2017.8203772}
\showDOI{\tempurl}


\bibitem[\protect\citeauthoryear{Ardeshiricham, Hu, Marxen, and
  Kastner}{Ardeshiricham et~al\mbox{.}}{2017b}]%
        {Armaiti2017}
\bibfield{author}{\bibinfo{person}{Armaiti Ardeshiricham}, \bibinfo{person}{Wei
  Hu}, \bibinfo{person}{Joshua Marxen}, {and} \bibinfo{person}{Ryan Kastner}.}
  \bibinfo{year}{2017}\natexlab{b}.
\newblock \showarticletitle{Register transfer level information flow tracking
  for provably secure hardware design}. In \bibinfo{booktitle}{\emph{Design,
  Automation Test in Europe Conference Exhibition (DATE), 2017}}.
  \bibinfo{pages}{1691--1696}.
\newblock
\urldef\tempurl%
\url{https://doi.org/10.23919/DATE.2017.7927266}
\showDOI{\tempurl}


\bibitem[\protect\citeauthoryear{Becker, Hu, Tai, Brisk, Kastner, and
  Ienne}{Becker et~al\mbox{.}}{2017}]%
        {Becker2017GLIFT}
\bibfield{author}{\bibinfo{person}{Andrew Becker}, \bibinfo{person}{Wei Hu},
  \bibinfo{person}{Yu Tai}, \bibinfo{person}{Philip Brisk},
  \bibinfo{person}{Ryan Kastner}, {and} \bibinfo{person}{Paolo Ienne}.}
  \bibinfo{year}{2017}\natexlab{}.
\newblock \showarticletitle{Arbitrary precision and complexity tradeoffs for
  gate-level information flow tracking}. In \bibinfo{booktitle}{\emph{2017 54th
  ACM/EDAC/IEEE Design Automation Conference (DAC)}}. \bibinfo{pages}{1--6}.
\newblock
\urldef\tempurl%
\url{https://doi.org/10.1145/3061639.3062203}
\showDOI{\tempurl}


\bibitem[\protect\citeauthoryear{Bidmeshki and Makris}{Bidmeshki and
  Makris}{2015}]%
        {bidmeshki2015toward}
\bibfield{author}{\bibinfo{person}{Mohammad-Mahdi Bidmeshki} {and}
  \bibinfo{person}{Yiorgos Makris}.} \bibinfo{year}{2015}\natexlab{}.
\newblock \showarticletitle{Toward automatic proof generation for information
  flow policies in third-party hardware IP}. In \bibinfo{booktitle}{\emph{2015
  IEEE International Symposium on Hardware Oriented Security and Trust
  (HOST)}}. IEEE, \bibinfo{pages}{163--168}.
\newblock


\bibitem[\protect\citeauthoryear{Bilzor, Huffmire, Irvine, and Levin}{Bilzor
  et~al\mbox{.}}{2011}]%
        {IrvineHOST2011}
\bibfield{author}{\bibinfo{person}{M. Bilzor}, \bibinfo{person}{T. Huffmire},
  \bibinfo{person}{C. Irvine}, {and} \bibinfo{person}{T. Levin}.}
  \bibinfo{year}{2011}\natexlab{}.
\newblock \showarticletitle{Security Checkers: Detecting processor malicious
  inclusions at runtime}. In \bibinfo{booktitle}{\emph{International Symposium
  on Hardware-Oriented Security and Trust (HOST)}}. \bibinfo{publisher}{IEEE},
  \bibinfo{pages}{34--39}.
\newblock
\urldef\tempurl%
\url{https://doi.org/10.1109/HST.2011.5954992}
\showDOI{\tempurl}


\bibitem[\protect\citeauthoryear{Brown}{Brown}{2017}]%
        {MichaelThesis}
\bibfield{author}{\bibinfo{person}{Michael Brown}.}
  \bibinfo{year}{2017}\natexlab{}.
\newblock \bibinfo{booktitle}{\emph{Cross-validation Processor
  Specifications}}.
\newblock \bibinfo{type}{Master's Thesis}. \bibinfo{institution}{University of
  North Carolina at Chapel Hill}.
\newblock


\bibitem[\protect\citeauthoryear{Chang and Wang}{Chang and Wang}{2010}]%
        {ChangASPDAC2010}
\bibfield{author}{\bibinfo{person}{Po-Hsien Chang} {and} \bibinfo{person}{Li~C
  Wang}.} \bibinfo{year}{2010}\natexlab{}.
\newblock \showarticletitle{Automatic assertion extraction via sequential data
  mining of simulation traces}. In \bibinfo{booktitle}{\emph{15th Asia and
  South Pacific Design Automation Conference (ASP-DAC)}}. IEEE,
  \bibinfo{pages}{607--612}.
\newblock


\bibitem[\protect\citeauthoryear{Clarkson and Schneider}{Clarkson and
  Schneider}{2010}]%
        {Clarkson2010}
\bibfield{author}{\bibinfo{person}{Michael~R. Clarkson} {and}
  \bibinfo{person}{Fred~B. Schneider}.} \bibinfo{year}{2010}\natexlab{}.
\newblock \showarticletitle{Hyperproperties}.
\newblock \bibinfo{journal}{\emph{J. Comput. Secur.}} \bibinfo{volume}{18},
  \bibinfo{number}{6} (\bibinfo{date}{Sept.} \bibinfo{year}{2010}),
  \bibinfo{pages}{1157--1210}.
\newblock
\showISSN{0926-227X}
\newblock
\shownote{\url{http://dl.acm.org/citation.cfm?id=1891823.1891830}.}


\bibitem[\protect\citeauthoryear{{Danese}, {Ghasempouri}, and
  {Pravadelli}}{{Danese} et~al\mbox{.}}{2015}]%
        {danese2015automaticdate}
\bibfield{author}{\bibinfo{person}{A. {Danese}}, \bibinfo{person}{T.
  {Ghasempouri}}, {and} \bibinfo{person}{G. {Pravadelli}}.}
  \bibinfo{year}{2015}\natexlab{}.
\newblock \showarticletitle{Automatic extraction of assertions from execution
  traces of behavioural models}. In \bibinfo{booktitle}{\emph{Design,
  Automation Test in Europe Conference Exhibition (DATE)}}.
  \bibinfo{pages}{67--72}.
\newblock
\urldef\tempurl%
\url{https://doi.org/10.7873/DATE.2015.0110}
\showDOI{\tempurl}


\bibitem[\protect\citeauthoryear{{Danese}, {Pravadelli}, and
  {Zandonà}}{{Danese} et~al\mbox{.}}{2016}]%
        {danese2015automatic}
\bibfield{author}{\bibinfo{person}{A. {Danese}}, \bibinfo{person}{G.
  {Pravadelli}}, {and} \bibinfo{person}{I. {Zandonà}}.}
  \bibinfo{year}{2016}\natexlab{}.
\newblock \showarticletitle{Automatic generation of power state machines
  through dynamic mining of temporal assertions}. In
  \bibinfo{booktitle}{\emph{Design, Automation Test in Europe Conference
  Exhibition (DATE)}}. \bibinfo{pages}{606--611}.
\newblock


\bibitem[\protect\citeauthoryear{{Danese}, {Riva}, and {Pravadelli}}{{Danese}
  et~al\mbox{.}}{2017}]%
        {danese2017team}
\bibfield{author}{\bibinfo{person}{A. {Danese}}, \bibinfo{person}{N.~D.
  {Riva}}, {and} \bibinfo{person}{G. {Pravadelli}}.}
  \bibinfo{year}{2017}\natexlab{}.
\newblock \showarticletitle{{A-TEAM}: Automatic template-based assertion
  miner}. In \bibinfo{booktitle}{\emph{54th Design Automation Conference
  (DAC)}}. \bibinfo{publisher}{ACM/EDAC/IEEE}, \bibinfo{pages}{1--6}.
\newblock
\urldef\tempurl%
\url{https://doi.org/10.1145/3061639.3062206}
\showDOI{\tempurl}


\bibitem[\protect\citeauthoryear{Dessouky, Gens, Haney, Persyn, Kanuparthi,
  Khattri, Fung, Sadeghi, and Rajendran}{Dessouky et~al\mbox{.}}{2019}]%
        {dessouky2019hardfails}
\bibfield{author}{\bibinfo{person}{Ghada Dessouky}, \bibinfo{person}{David
  Gens}, \bibinfo{person}{Patrick Haney}, \bibinfo{person}{Garrett Persyn},
  \bibinfo{person}{Arun Kanuparthi}, \bibinfo{person}{Hareesh Khattri},
  \bibinfo{person}{Jason~M Fung}, \bibinfo{person}{Ahmad-Reza Sadeghi}, {and}
  \bibinfo{person}{Jeyavijayan Rajendran}.} \bibinfo{year}{2019}\natexlab{}.
\newblock \showarticletitle{Hardfails: Insights into Software-Exploitable
  Hardware Bugs}. In \bibinfo{booktitle}{\emph{28th USENIX Security
  Symposium}}. \bibinfo{publisher}{USENIX Association},
  \bibinfo{pages}{213--230}.
\newblock
\urldef\tempurl%
\url{https://www.usenix.org/conference/usenixsecurity19/presentation/dessouky}
\showURL{%
\tempurl}


\bibitem[\protect\citeauthoryear{Deutschbein and Sturton}{Deutschbein and
  Sturton}{2018}]%
        {deutschbein2018mining}
\bibfield{author}{\bibinfo{person}{Calvin Deutschbein} {and}
  \bibinfo{person}{Cynthia Sturton}.} \bibinfo{year}{2018}\natexlab{}.
\newblock \showarticletitle{Mining Security Critical Linear Temporal Logic
  Specifications for Processors}. In \bibinfo{booktitle}{\emph{International
  Workshop on Microprocessor and SoC Test, Security, and Verification (MTV)}}.
  \bibinfo{publisher}{IEEE}.
\newblock
\urldef\tempurl%
\url{https://ieeexplore.ieee.org/document/8746060}
\showURL{%
\tempurl}


\bibitem[\protect\citeauthoryear{{Deutschbein} and {Sturton}}{{Deutschbein} and
  {Sturton}}{2020}]%
        {deutschbein2020CISC}
\bibfield{author}{\bibinfo{person}{C. {Deutschbein}} {and} \bibinfo{person}{C.
  {Sturton}}.} \bibinfo{year}{2020}\natexlab{}.
\newblock \showarticletitle{Evaluating Security Specification Mining for a CISC
  Architecture}. In \bibinfo{booktitle}{\emph{2020 IEEE International Symposium
  on Hardware Oriented Security and Trust (HOST)}}. \bibinfo{pages}{164--175}.
\newblock
\urldef\tempurl%
\url{https://doi.org/10.1109/HOST45689.2020.9300291}
\showDOI{\tempurl}


\bibitem[\protect\citeauthoryear{Deutschbein and Sturton}{Deutschbein and
  Sturton}{2020}]%
        {deutschbein2020evaluatingshort}
\bibfield{author}{\bibinfo{person}{Calvin Deutschbein} {and}
  \bibinfo{person}{Cynthia Sturton}.} \bibinfo{year}{2020}\natexlab{}.
\newblock \showarticletitle{Evaluating Security Specification Mining for a
  {CISC} Architecture}. In \bibinfo{booktitle}{\emph{Symposium on Hardware
  Oriented Security and Trust (HOST)}}. \bibinfo{publisher}{IEEE}.
\newblock


\bibitem[\protect\citeauthoryear{Ernst, Perkins, Guo, McCamant, Pacheco,
  Tschantz, and Xiao}{Ernst et~al\mbox{.}}{2007}]%
        {Ernst2007}
\bibfield{author}{\bibinfo{person}{Michael~D. Ernst}, \bibinfo{person}{Jeff~H.
  Perkins}, \bibinfo{person}{Philip~J. Guo}, \bibinfo{person}{Stephen
  McCamant}, \bibinfo{person}{Carlos Pacheco}, \bibinfo{person}{Matthew~S.
  Tschantz}, {and} \bibinfo{person}{Chen Xiao}.}
  \bibinfo{year}{2007}\natexlab{}.
\newblock \showarticletitle{The {Daikon} System for Dynamic Detection of Likely
  Invariants}.
\newblock \bibinfo{journal}{\emph{Science of Computer Programming}}
  \bibinfo{volume}{69}, \bibinfo{number}{1-3} (\bibinfo{date}{Dec.}
  \bibinfo{year}{2007}), \bibinfo{pages}{35--45}.
\newblock
\showISSN{0167-6423}
\urldef\tempurl%
\url{https://doi.org/10.1016/j.scico.2007.01.015}
\showDOI{\tempurl}
\newblock
\shownote{\url{http://dx.doi.org/10.1016/j.scico.2007.01.015}.}


\bibitem[\protect\citeauthoryear{Farzana, Rahman, Tehranipoor, and
  Farahmandi}{Farzana et~al\mbox{.}}{2019}]%
        {Farzana2019SoC}
\bibfield{author}{\bibinfo{person}{Nusrat Farzana}, \bibinfo{person}{Fahim
  Rahman}, \bibinfo{person}{Mark Tehranipoor}, {and} \bibinfo{person}{Farimah
  Farahmandi}.} \bibinfo{year}{2019}\natexlab{}.
\newblock \showarticletitle{SoC Security Verification using Property Checking}.
  In \bibinfo{booktitle}{\emph{2019 IEEE International Test Conference (ITC)}}.
  \bibinfo{pages}{1--10}.
\newblock
\urldef\tempurl%
\url{https://doi.org/10.1109/ITC44170.2019.9000170}
\showDOI{\tempurl}


\bibitem[\protect\citeauthoryear{Gabel and Su}{Gabel and Su}{2008a}]%
        {gabel2008javert}
\bibfield{author}{\bibinfo{person}{Mark Gabel} {and} \bibinfo{person}{Zhendong
  Su}.} \bibinfo{year}{2008}\natexlab{a}.
\newblock \showarticletitle{Javert: Fully Automatic Mining of General Temporal
  Properties from Dynamic Traces}. In \bibinfo{booktitle}{\emph{16th
  International Symposium on Foundations of Software Engineering (FSE)}}.
  \bibinfo{publisher}{ACM}, \bibinfo{pages}{339--349}.
\newblock
\showISBNx{978-1-59593-995-1}
\urldef\tempurl%
\url{https://doi.org/10.1145/1453101.1453150}
\showDOI{\tempurl}
\newblock
\shownote{\url{http://doi.acm.org/10.1145/1453101.1453150}.}


\bibitem[\protect\citeauthoryear{Gabel and Su}{Gabel and Su}{2008b}]%
        {gabel2008symbolic}
\bibfield{author}{\bibinfo{person}{Mark Gabel} {and} \bibinfo{person}{Zhendong
  Su}.} \bibinfo{year}{2008}\natexlab{b}.
\newblock \showarticletitle{Symbolic Mining of Temporal Specifications}. In
  \bibinfo{booktitle}{\emph{30th International Conference on Software
  Engineering (ICSE)}}. \bibinfo{publisher}{ACM}, \bibinfo{pages}{51--60}.
\newblock
\showISBNx{978-1-60558-079-1}
\urldef\tempurl%
\url{https://doi.org/10.1145/1368088.1368096}
\showDOI{\tempurl}
\newblock
\shownote{\url{http://doi.acm.org/10.1145/1368088.1368096}.}


\bibitem[\protect\citeauthoryear{Hangal, Chandra, Narayanan, and
  Chakravorty}{Hangal et~al\mbox{.}}{2005a}]%
        {hangal2005iodine}
\bibfield{author}{\bibinfo{person}{Sudheendra Hangal}, \bibinfo{person}{Naveen
  Chandra}, \bibinfo{person}{Sridhar Narayanan}, {and} \bibinfo{person}{Sandeep
  Chakravorty}.} \bibinfo{year}{2005}\natexlab{a}.
\newblock \showarticletitle{{IODINE}: a tool to automatically infer dynamic
  invariants for hardware designs}. In \bibinfo{booktitle}{\emph{42nd annual
  Design Automation Conference}}. ACM, \bibinfo{pages}{775--778}.
\newblock
\urldef\tempurl%
\url{http://xenon.stanford.edu/~hangal/iodine.html}
\showURL{%
\tempurl}


\bibitem[\protect\citeauthoryear{Hangal, Narayanan, Chandra, and
  Chakravorty}{Hangal et~al\mbox{.}}{2005b}]%
        {HangalDAC2005}
\bibfield{author}{\bibinfo{person}{Sudheendra Hangal}, \bibinfo{person}{Sridhar
  Narayanan}, \bibinfo{person}{Naveen Chandra}, {and} \bibinfo{person}{Sandeep
  Chakravorty}.} \bibinfo{year}{2005}\natexlab{b}.
\newblock \showarticletitle{{IODINE}: A tool to automatically infer dynamic
  invariants for hardware designs}. In \bibinfo{booktitle}{\emph{42nd Design
  Automation Conference (DAC)}}. \bibinfo{publisher}{IEEE}.
\newblock


\bibitem[\protect\citeauthoryear{Hertz, Sheridan, and Vasudevan}{Hertz
  et~al\mbox{.}}{2013}]%
        {HertzTCAD2013}
\bibfield{author}{\bibinfo{person}{Stav Hertz}, \bibinfo{person}{David
  Sheridan}, {and} \bibinfo{person}{Shobha Vasudevan}.}
  \bibinfo{year}{2013}\natexlab{}.
\newblock \showarticletitle{Mining hardware assertions with guidance from
  static analysis}.
\newblock \bibinfo{journal}{\emph{Transactions on Computer-Aided Design of
  Integrated Circuits and Systems}} \bibinfo{volume}{32}, \bibinfo{number}{6}
  (\bibinfo{year}{2013}), \bibinfo{pages}{952--965}.
\newblock


\bibitem[\protect\citeauthoryear{Hicks, Sturton, King, and Smith}{Hicks
  et~al\mbox{.}}{2015}]%
        {HicksASPLOS2015}
\bibfield{author}{\bibinfo{person}{Matthew Hicks}, \bibinfo{person}{Cynthia
  Sturton}, \bibinfo{person}{Samuel~T. King}, {and}
  \bibinfo{person}{Jonathan~M. Smith}.} \bibinfo{year}{2015}\natexlab{}.
\newblock \showarticletitle{{SPECS}: A Lightweight Runtime Mechanism for
  Protecting Software from Security-Critical Processor Bugs}. In
  \bibinfo{booktitle}{\emph{Twentieth International Conference on Architectural
  Support for Programming Languages and Operating Systems (ASPLOS)}}.
  \bibinfo{publisher}{ACM}, \bibinfo{pages}{517--529}.
\newblock
\showISBNx{978-1-4503-2835-7}
\urldef\tempurl%
\url{https://doi.org/10.1145/2694344.2694366}
\showDOI{\tempurl}
\newblock
\shownote{\url{http://doi.acm.org/10.1145/2694344.2694366}.}


\bibitem[\protect\citeauthoryear{Hu, Althoff, Ardeshiricham, and Kastner}{Hu
  et~al\mbox{.}}{2016}]%
        {hu2016towards}
\bibfield{author}{\bibinfo{person}{Wei Hu}, \bibinfo{person}{Alric Althoff},
  \bibinfo{person}{Armaiti Ardeshiricham}, {and} \bibinfo{person}{Ryan
  Kastner}.} \bibinfo{year}{2016}\natexlab{}.
\newblock \showarticletitle{Towards property driven hardware security}. In
  \bibinfo{booktitle}{\emph{2016 17th International Workshop on Microprocessor
  and SOC Test and Verification (MTV)}}. IEEE, \bibinfo{pages}{51--56}.
\newblock


\bibitem[\protect\citeauthoryear{Hu, Ardeshiricham, Gobulukoglu, Wang, and
  Kastner}{Hu et~al\mbox{.}}{2018}]%
        {Armaiti2018}
\bibfield{author}{\bibinfo{person}{Wei Hu}, \bibinfo{person}{Armaiti
  Ardeshiricham}, \bibinfo{person}{Mustafa~S Gobulukoglu},
  \bibinfo{person}{Xinmu Wang}, {and} \bibinfo{person}{Ryan Kastner}.}
  \bibinfo{year}{2018}\natexlab{}.
\newblock \showarticletitle{Property Specific Information Flow Analysis for
  Hardware Security Verification} \emph{(\bibinfo{series}{ICCAD '18})}.
  \bibinfo{publisher}{ACM}.
\newblock
\showISBNx{9781450359504}
\urldef\tempurl%
\url{https://doi.org/10.1145/3240765.3240839}
\showDOI{\tempurl}


\bibitem[\protect\citeauthoryear{Hu, Mu, Oberg, Mao, Tiwari, Sherwood, and
  Kastner}{Hu et~al\mbox{.}}{2014}]%
        {hu2014glift}
\bibfield{author}{\bibinfo{person}{Wei Hu}, \bibinfo{person}{Dejun Mu},
  \bibinfo{person}{Jason Oberg}, \bibinfo{person}{Baolei Mao},
  \bibinfo{person}{Mohit Tiwari}, \bibinfo{person}{Timothy Sherwood}, {and}
  \bibinfo{person}{Ryan Kastner}.} \bibinfo{year}{2014}\natexlab{}.
\newblock \showarticletitle{Gate-Level Information Flow Tracking for Security
  Lattices}.
\newblock \bibinfo{journal}{\emph{ACM Trans. Des. Autom. Electron. Syst.}}
  \bibinfo{volume}{20}, \bibinfo{number}{1}, Article \bibinfo{articleno}{2}
  (\bibinfo{date}{Nov.} \bibinfo{year}{2014}), \bibinfo{numpages}{25}~pages.
\newblock
\showISSN{1084-4309}
\urldef\tempurl%
\url{https://doi.org/10.1145/2676548}
\showDOI{\tempurl}
\newblock
\shownote{\url{https://doi.org/10.1145/2676548}.}


\bibitem[\protect\citeauthoryear{Li, Forin, and Seshia}{Li
  et~al\mbox{.}}{2010}]%
        {li2010scalable}
\bibfield{author}{\bibinfo{person}{Wenchao Li}, \bibinfo{person}{Alessandro
  Forin}, {and} \bibinfo{person}{Sanjit~A. Seshia}.}
  \bibinfo{year}{2010}\natexlab{}.
\newblock \showarticletitle{Scalable Specification Mining for Verification and
  Diagnosis}. In \bibinfo{booktitle}{\emph{47th Design Automation Conference
  (DAC)}}. \bibinfo{publisher}{ACM}, \bibinfo{pages}{755--760}.
\newblock
\newblock
\shownote{\url{http://doi.acm.org/10.1145/1837274.1837466}.}


\bibitem[\protect\citeauthoryear{{Liu}, {Lin}, and {Vasudevan}}{{Liu}
  et~al\mbox{.}}{2012}]%
        {liu2012word}
\bibfield{author}{\bibinfo{person}{L. {Liu}}, \bibinfo{person}{C. {Lin}}, {and}
  \bibinfo{person}{S. {Vasudevan}}.} \bibinfo{year}{2012}\natexlab{}.
\newblock \showarticletitle{Word level feature discovery to enhance quality of
  assertion mining}. In \bibinfo{booktitle}{\emph{International Conference on
  Computer-Aided Design (ICCAD)}}. \bibinfo{publisher}{IEEE/ACM},
  \bibinfo{pages}{210--217}.
\newblock


\bibitem[\protect\citeauthoryear{Meng, Kundu, Kanuparthi, and Basu}{Meng
  et~al\mbox{.}}{2021}]%
        {Meng2021Concolic}
\bibfield{author}{\bibinfo{person}{Xingyu Meng}, \bibinfo{person}{Shamik
  Kundu}, \bibinfo{person}{Arun~K. Kanuparthi}, {and} \bibinfo{person}{Kanad
  Basu}.} \bibinfo{year}{2021}\natexlab{}.
\newblock \showarticletitle{RTL-ConTest: Concolic Testing on RTL for Detecting
  Security Vulnerabilities}.
\newblock \bibinfo{journal}{\emph{IEEE Transactions on Computer-Aided Design of
  Integrated Circuits and Systems}} (\bibinfo{year}{2021}),
  \bibinfo{pages}{1--1}.
\newblock
\urldef\tempurl%
\url{https://doi.org/10.1109/TCAD.2021.3066560}
\showDOI{\tempurl}


\bibitem[\protect\citeauthoryear{Min, Kashyap, Lee, Song, and Kim}{Min
  et~al\mbox{.}}{2015}]%
        {MinSOSP2015}
\bibfield{author}{\bibinfo{person}{Changwoo Min}, \bibinfo{person}{Sanidhya
  Kashyap}, \bibinfo{person}{Byoungyoung Lee}, \bibinfo{person}{Chengyu Song},
  {and} \bibinfo{person}{Taesoo Kim}.} \bibinfo{year}{2015}\natexlab{}.
\newblock \showarticletitle{Cross-checking Semantic Correctness: The Case of
  Finding File System Bugs}. In \bibinfo{booktitle}{\emph{25th Symposium on
  Operating Systems Principles (SOSP)}}. \bibinfo{publisher}{ACM},
  \bibinfo{pages}{361--377}.
\newblock
\showISBNx{978-1-4503-3834-9}
\urldef\tempurl%
\url{https://doi.org/10.1145/2815400.2815422}
\showDOI{\tempurl}
\newblock
\shownote{\url{http://doi.acm.org/10.1145/2815400.2815422}.}


\bibitem[\protect\citeauthoryear{Perkins, Kim, Larsen, Amarasinghe, Bachrach,
  Carbin, Pacheco, Sherwood, Sidiroglou, Sullivan, Wong, Zibin, Ernst, and
  Rinard}{Perkins et~al\mbox{.}}{2009}]%
        {PerkinsSOSP2009}
\bibfield{author}{\bibinfo{person}{Jeff~H. Perkins}, \bibinfo{person}{Sunghun
  Kim}, \bibinfo{person}{Sam Larsen}, \bibinfo{person}{Saman Amarasinghe},
  \bibinfo{person}{Jonathan Bachrach}, \bibinfo{person}{Michael Carbin},
  \bibinfo{person}{Carlos Pacheco}, \bibinfo{person}{Frank Sherwood},
  \bibinfo{person}{Stelios Sidiroglou}, \bibinfo{person}{Greg Sullivan},
  \bibinfo{person}{Weng-Fai Wong}, \bibinfo{person}{Yoav Zibin},
  \bibinfo{person}{Michael~D. Ernst}, {and} \bibinfo{person}{Martin Rinard}.}
  \bibinfo{year}{2009}\natexlab{}.
\newblock \showarticletitle{Automatically Patching Errors in Deployed
  Software}. In \bibinfo{booktitle}{\emph{22nd Symposium on Operating Systems
  Principles (SOSP)}}. \bibinfo{publisher}{ACM}, \bibinfo{pages}{87--102}.
\newblock
\showISBNx{978-1-60558-752-3}
\urldef\tempurl%
\url{https://doi.org/10.1145/1629575.1629585}
\showDOI{\tempurl}
\newblock
\shownote{\url{http://doi.acm.org/10.1145/1629575.1629585}.}


\bibitem[\protect\citeauthoryear{Pilato, Wu, Garg, Karri, and Regazzoni}{Pilato
  et~al\mbox{.}}{2019}]%
        {pilato2019tainthls}
\bibfield{author}{\bibinfo{person}{Christian Pilato}, \bibinfo{person}{Kaijie
  Wu}, \bibinfo{person}{Siddharth Garg}, \bibinfo{person}{Ramesh Karri}, {and}
  \bibinfo{person}{Francesco Regazzoni}.} \bibinfo{year}{2019}\natexlab{}.
\newblock \showarticletitle{TaintHLS: High-Level Synthesis for Dynamic
  Information Flow Tracking}.
\newblock \bibinfo{journal}{\emph{IEEE Transactions on Computer-Aided Design of
  Integrated Circuits and Systems}} \bibinfo{volume}{38}, \bibinfo{number}{5}
  (\bibinfo{year}{2019}), \bibinfo{pages}{798--808}.
\newblock
\urldef\tempurl%
\url{https://doi.org/10.1109/TCAD.2018.2834421}
\showDOI{\tempurl}


\bibitem[\protect\citeauthoryear{Rawat, Muduli, and Subramanyan}{Rawat
  et~al\mbox{.}}{2020}]%
        {Rawat2020Hyperminer}
\bibfield{author}{\bibinfo{person}{Mayank Rawat}, \bibinfo{person}{Sujit~Kumar
  Muduli}, {and} \bibinfo{person}{Pramod Subramanyan}.}
  \bibinfo{year}{2020}\natexlab{}.
\newblock \showarticletitle{Mining Hyperproperties from Behavioral Traces}. In
  \bibinfo{booktitle}{\emph{2020 IFIP/IEEE 28th International Conference on
  Very Large Scale Integration (VLSI-SOC)}}. \bibinfo{pages}{88--93}.
\newblock
\urldef\tempurl%
\url{https://doi.org/10.1109/VLSI-SOC46417.2020.9344106}
\showDOI{\tempurl}


\bibitem[\protect\citeauthoryear{{Reger}, {Barringer}, and {Rydeheard}}{{Reger}
  et~al\mbox{.}}{2013}]%
        {reger2013pattern}
\bibfield{author}{\bibinfo{person}{G. {Reger}}, \bibinfo{person}{H.
  {Barringer}}, {and} \bibinfo{person}{D. {Rydeheard}}.}
  \bibinfo{year}{2013}\natexlab{}.
\newblock \showarticletitle{A pattern-based approach to parametric
  specification mining}. In \bibinfo{booktitle}{\emph{28th International
  Conference on Automated Software Engineering (ASE)}}.
  \bibinfo{publisher}{IEEE/ACM}, \bibinfo{pages}{658--663}.
\newblock
\urldef\tempurl%
\url{https://doi.org/10.1109/ASE.2013.6693129}
\showDOI{\tempurl}


\bibitem[\protect\citeauthoryear{Restuccia, Meza, and Kastner}{Restuccia
  et~al\mbox{.}}{2021}]%
        {Restuccia2021AKER}
\bibfield{author}{\bibinfo{person}{Francesco Restuccia},
  \bibinfo{person}{Andres Meza}, {and} \bibinfo{person}{Ryan Kastner}.}
  \bibinfo{year}{2021}\natexlab{}.
\newblock \showarticletitle{{AKER:} {A} Design and Verification Framework for
  Safe and Secure SoC Access Control}.
\newblock \bibinfo{journal}{\emph{CoRR}}  \bibinfo{volume}{abs/2106.13263}
  (\bibinfo{year}{2021}).
\newblock
\showeprint[arxiv]{2106.13263}
\urldef\tempurl%
\url{https://arxiv.org/abs/2106.13263}
\showURL{%
\tempurl}


\bibitem[\protect\citeauthoryear{Tan, Zhang, Ma, Xiong, and Zhou}{Tan
  et~al\mbox{.}}{2008}]%
        {TanUSENIXSec2008}
\bibfield{author}{\bibinfo{person}{Lin Tan}, \bibinfo{person}{Xiaolan Zhang},
  \bibinfo{person}{Xiao Ma}, \bibinfo{person}{Weiwei Xiong}, {and}
  \bibinfo{person}{Yuanyuan Zhou}.} \bibinfo{year}{2008}\natexlab{}.
\newblock \showarticletitle{{AutoISES}: Automatically Inferring Security
  Specifications and Detecting Violations}. In \bibinfo{booktitle}{\emph{17th
  USENIX Security Symposium}}. \bibinfo{publisher}{USENIX Association},
  \bibinfo{pages}{379--394}.
\newblock
\newblock
\shownote{\url{http://dl.acm.org/citation.cfm?id=1496711.1496737}.}


\bibitem[\protect\citeauthoryear{{Wei Hu}, {Becker}, {Ardeshiricham}, {Yu Tai},
  {Ienne}, {Mu}, and {Kastner}}{{Wei Hu} et~al\mbox{.}}{2016}]%
        {Hu2016}
\bibfield{author}{\bibinfo{person}{{Wei Hu}}, \bibinfo{person}{A. {Becker}},
  \bibinfo{person}{A. {Ardeshiricham}}, \bibinfo{person}{{Yu Tai}},
  \bibinfo{person}{P. {Ienne}}, \bibinfo{person}{D. {Mu}}, {and}
  \bibinfo{person}{R. {Kastner}}.} \bibinfo{year}{2016}\natexlab{}.
\newblock \showarticletitle{Imprecise security: Quality and complexity
  tradeoffs for hardware information flow tracking}. In
  \bibinfo{booktitle}{\emph{2016 IEEE/ACM International Conference on
  Computer-Aided Design (ICCAD)}}. \bibinfo{pages}{1--8}.
\newblock
\showISSN{1558-2434}
\urldef\tempurl%
\url{https://doi.org/10.1145/2966986.2967046}
\showDOI{\tempurl}


\bibitem[\protect\citeauthoryear{Weimer and Necula}{Weimer and Necula}{2005}]%
        {weimer2005mining}
\bibfield{author}{\bibinfo{person}{Westley Weimer} {and}
  \bibinfo{person}{George~C. Necula}.} \bibinfo{year}{2005}\natexlab{}.
\newblock \showarticletitle{Mining Temporal Specifications for Error
  Detection}. In \bibinfo{booktitle}{\emph{11th International Conference on
  Tools and Algorithms for the Construction and Analysis of Systems (TACAS)}}.
  \bibinfo{publisher}{Springer-Verlag}, \bibinfo{pages}{461--476}.
\newblock
\showISBNx{3-540-25333-5, 978-3-540-25333-4}
\urldef\tempurl%
\url{https://doi.org/10.1007/978-3-540-31980-1_30}
\showDOI{\tempurl}
\newblock
\shownote{\url{http://dx.doi.org/10.1007/978-3-540-31980-1_30}.}


\bibitem[\protect\citeauthoryear{Yamaguchi, Lindner, and Rieck}{Yamaguchi
  et~al\mbox{.}}{2011}]%
        {YamaguchiWOOT2011}
\bibfield{author}{\bibinfo{person}{Fabian Yamaguchi}, \bibinfo{person}{Felix
  Lindner}, {and} \bibinfo{person}{Konrad Rieck}.}
  \bibinfo{year}{2011}\natexlab{}.
\newblock \showarticletitle{Vulnerability Extrapolation: Assisted Discovery of
  Vulnerabilities Using Machine Learning}. In \bibinfo{booktitle}{\emph{5th
  USENIX Conference on Offensive Technologies (WOOT)}}.
  \bibinfo{publisher}{USENIX Association}, \bibinfo{pages}{13--13}.
\newblock
\newblock
\shownote{\url{http://dl.acm.org/citation.cfm?id=2028052.2028065}.}


\bibitem[\protect\citeauthoryear{Yang, Evans, Bhardwaj, Bhat, and Das}{Yang
  et~al\mbox{.}}{2006}]%
        {yang2006perracotta}
\bibfield{author}{\bibinfo{person}{Jinlin Yang}, \bibinfo{person}{David Evans},
  \bibinfo{person}{Deepali Bhardwaj}, \bibinfo{person}{Thirumalesh Bhat}, {and}
  \bibinfo{person}{Manuvir Das}.} \bibinfo{year}{2006}\natexlab{}.
\newblock \showarticletitle{Perracotta: Mining Temporal {API} Rules from
  Imperfect Traces}. In \bibinfo{booktitle}{\emph{28th International Conference
  on Software Engineering (ICSE)}}. \bibinfo{publisher}{ACM},
  \bibinfo{pages}{282--291}.
\newblock
\showISBNx{1-59593-375-1}
\urldef\tempurl%
\url{https://doi.org/10.1145/1134285.1134325}
\showDOI{\tempurl}
\newblock
\shownote{\url{http://doi.acm.org/10.1145/1134285.1134325}.}


\bibitem[\protect\citeauthoryear{Zhang, Stanley, Griggs, Chi, and
  Sturton}{Zhang et~al\mbox{.}}{2017}]%
        {zhang2017identifyingshort}
\bibfield{author}{\bibinfo{person}{Rui Zhang}, \bibinfo{person}{Natalie
  Stanley}, \bibinfo{person}{Chris Griggs}, \bibinfo{person}{Andrew Chi}, {and}
  \bibinfo{person}{Cynthia Sturton}.} \bibinfo{year}{2017}\natexlab{}.
\newblock \showarticletitle{Identifying Security Critical Properties for the
  Dynamic Verification of a Processor}. In
  \bibinfo{booktitle}{\emph{Architectural Support for Prog. Lang. and Operating
  Sys. (ASPLOS)}}. \bibinfo{publisher}{ACM}.
\newblock


\end{thebibliography}

\appendix
\begin{appendix}




\pagebreak

\section{Sample Properties}
\label{sec:samples}

In this section we show examples of Isadora output.

\subsection{Case 154: ACW Security Property}
\label{sec:sampleoutput}

\floatstyle{boxed}
\restylefloat{figure}

\begin{figure}[h]
    \begin{verbatim}
case 154: 2_121_250_379_543
	_src_ in {w_base_addr_wire, M_AXI_AWREADY_wire, 
	AW_CH_DIS, 	w_max_outs_wire, AW_ILLEGAL_REQ, 
	w_num_trans_wire, AW_STATE, 	AW_CH_EN}  
	=/=> 
	_snk_ in {M_AXI_WDATA}
	unless
0 != _inv_ in {ADDR_LSB, ARESETN, M_AXI_ARBURST_wire, 
M_AXI_ARCACHE_wire, M_AXI_ARLEN_wire, M_AXI_ARREADY, 
M_AXI_ARSIZE_wire, M_AXI_AWBURST_wire, 
M_AXI_AWCACHE_wire, M_AXI_AWLEN_wire, M_AXI_AWREADY, 
M_AXI_AWSIZE_wire, M_AXI_BREADY, M_AXI_BREADY_wire, 
M_AXI_WREADY, M_AXI_WREADY_wire, M_AXI_WSTRB_wire, 
OPT_MEM_ADDR_BITS, S_AXI_CTRL_BREADY, 
S_AXI_CTRL_RREADY, data_val_wire, r_burst_len_wire, 
r_displ_wire, r_max_outs_wire, r_num_trans_wire, 
r_phase_wire, w_burst_len_wire, w_displ_wire, 
w_max_outs_wire, w_num_trans_wire, w_phase_wire}
    \end{verbatim}
    \caption{An example of an Isadora property, Case 154, over the Single ACW}
    \label{fig:isaprop}
\end{figure}
\floatstyle{plain}
\restylefloat{figure}

To consider the output properties of Isadora, Figures~\ref{fig:isaprop}
shows an example of Isadora output, Case 154 of the 303 output
properties over the ACW module. 
This a case that was sampled during evaluation.
Here the condition predicates shown are register equality testing versus
zero. Other predicates are
captured within the workflow but not propagated to individual properties
formatted for output.

A visible difference between an Isadora output property 
and the property grammar of Section~\ref{sec:props} is that 
at output stage Isadora properties may specify multiple source registers,
may consider multiple sink registers though do not do so in
this case, and may contain multiple invariants as conditions.

Case 154 includes an example of a flow condition
between internal and peripheral visible signals in addition to specifying
other aspects of design behavior. This is similar to the
example of write readiness from Section~\ref{sec:props}, 
but in Case 154, the flow is from
the internal signal to the peripheral, though the power state 
predicate is identical.
Of note, as in the case of write readiness,
this flow occurs exclusively within the write channel, as denoted
by the ``$\mathtt{W}$'' present in ready wire and the data register.
\begin{center}
 $\mathtt{AWREADY\_int}$ \texttt{=/=>}
 $\mathtt{WDATA}$ unless $(\mathtt{ARESETN}$ $\neq 0)$
 \end{center}

\subsubsection{Security Relevance}

Under the working definition of security properties for Isadora,
where internal signals and peripheral signals should not flow to
one another unless ACW is not undergoing a reset, this
single source, single sink, single invariant description of behavior
composed from an Isadora output property establishes Case 154
as a security property under the working definition.
Case 154 describes signals marked as sensitive by designers,
both labeled as such within the design using comments
and present within security properties they specified, and
differs from a designer provided property only in the specific
pairing of registers.

\subsection{Case 144: ACW Functional Property}
\label{sec:case144}

One example of an Isadora property classified as functional,
with truncated flow conditions, is presented in Figure~\ref{fig:isafunc},
and captures a logical update to an internal decoder signal.
This additional shows an example of a property over multiple sinks,
a single source, and for which there are predicates capturing
both equality and inequality to zero.

\floatstyle{boxed}
\restylefloat{figure}

\begin{figure}[h]
    \begin{verbatim}
case 144: 128
	_src_ in {instr_lw}  
	=/=> 
	_snk_ in {is_slti_blt_slt, is_sltiu_bltu_sltu}
	unless
0 == _r_ in {alu_eq, alu_shl, alu_shr,  ... }
0 != _r_ in {alu_add_sub, alu_lts, alu_ltu, ... }\end{verbatim}
    \caption{An example of an Isadora property, Case 144, over RISC-V.}
    \label{fig:isafunc}
\end{figure}
\floatstyle{plain}
\restylefloat{figure}

\end{appendix}

\end{document}